\documentclass{elsart}

\usepackage{graphicx}
\usepackage{epstopdf}
\usepackage{amssymb}

\begin{document}

\begin{frontmatter}

% Title, authors and addresses

% use the thanksref command within \title, \author or \address for footnotes;
% use the corauthref command within \author for corresponding author footnotes;
% use the ead command for the email address,
% and the form \ead[url] for the home page:
% \title{Title\thanksref{label1}}
% \thanks[label1]{}
% \author{Name\corauthref{cor1}\thanksref{label2}}
% \ead{email address}
% \ead[url]{home page}
% \thanks[label2]{}
% \corauth[cor1]{}
% \address{Address\thanksref{label3}}
% \thanks[label3]{}

\title{Correlations of Structure and Dynamics in an Aging Colloidal Glass}

% use optional labels to link authors explicitly to addresses:
% \author[label1,label2]{}
% \address[label1]{}
% \address[label2]{}

\author{Gianguido C. Cianci, Rachel E. Courtland and Eric R. Weeks\corauthref{cor1}}
\corauth[cor1]{For correspondence: weeks@physics.emory.edu}
\address{Department of Physics, Mail Stop 1131/002/1AB,
Emory University, Atlanta, GA 30322, U.S.A.}

\begin{abstract}
We study concentrated colloidal suspensions, a model system
which has a glass transition.  Samples in the glassy state show
aging, in that the motion of the colloidal particles slows as the
sample ages from an initial state.  We study the relationship
between the static structure and the slowing dynamics, using
confocal microscopy to follow the three-dimensional motion of the
particles.  The structure is quantified by considering tetrahedra
formed by quadruplets of neighboring particles.  We
find that while the sample clearly slows down during aging,
the static properties as measured by tetrahedral quantities
do not vary. However, a weak correlation between tetrahedron
shape and mobility is observed, suggesting that the
structure facilitates the motion responsible for the sample aging.
\end{abstract}

\end{frontmatter}

% main text
\section{Introduction}

When some liquids undergo a rapid temperature quench they can
form glasses.  This occurs at a glass transition temperature
$T_g$ which often depends on the cooling rate.  As the system
is cooled, the approaching glass transition is marked by a
dramatic increase in the macroscopic viscosity of the liquid,
and a corresponding increase in the microscopic time scales
for motion \cite{review1,review2,review3,review4}.  Both the
viscosity and the microscopic relaxation time can change by many
orders of magnitude as the temperature decreases by merely 10\%.
Once in the glass state, another phenomenon is noted, that of aging:
the dependence of the properties of the system on the time
elapsed since reaching $T_g$.  When such behavior is observed
the system is said to be out of equilibrium, a fact that could
be anticipated by noting that the dependence of $T_g$
itself on cooling rate implies the glass transition is not
an equilibrium phenomenon.  Aging most prominently manifests
itself in the dynamics:  the microscopic relaxation time scale
depends on the age of the sample.

Attempts to explain these phenomena try to link the microscopic
structure to the microscopic dynamics.  For example, one might
postulate that the increase in viscosity is caused by the growth
of domains whose dynamics are correlated \cite{ediger00}.
However, no experiment has seen a structural length scale
characterizing such domains that grows or diverges at $T_g$
\cite{review1,menon94,blaaderen95}.  Likewise, aging might be due to some
coarsening of structure; as domains of glassy structure grow,
motion is slowed, which in turn slows the further growth of
these domains.  However, these domains have not been identified,
and currently no structural features have been identified that
explain aging dynamics \cite{vanmegen98,kob00}.

Recently, some interesting developments in the
study of aging non-equilibrium systems have been
brought about by the adoption of dense colloidal
suspensions as model systems for liquids, glasses and
gels
\cite{blaaderen95,vanmegen98,pusey86,cipelletti00,cipelletti03,courtland}.
Colloidal suspensions consist of solid particles in a liquid,
and the motion of the particles is analogous to that of atoms
or molecules in a more traditional material
\cite{pusey86,kegel00,weeks00}.  In these systems,
the particle interactions can easily be tuned from repulsive
to attractive.  A common case is when particles interact simply
as hard spheres with no interactions, attractive or
repulsive, other than when they are in direct contact
\cite{pusey86,snook91}.  In all cases, a major control parameter
is the packing fraction $\phi$. For hard spheres this is the
only control parameter and when $\phi$ is raised above a value
of $\phi_{\rm g}\approx 0.58$ the system becomes
glassy and the aging process begins.

While structural changes in an aging system remain unclear, two
experiments studying colloids have characterized the dynamics.
Cipelletti and co-workers \cite{cipelletti03} studied aging
in a colloidal gel using novel light scattering techniques
and showed that the dynamics in such a non-equilibrium system
present striking temporal heterogeneities.  Aging has also
been studied in a colloidal glass using confocal microscopy
\cite{courtland}. In that study both temporal and spatial
heterogeneities were seen.  However, despite the ease with
which a colloidal glass can be formed and observed a detailed
understanding of the structural changes that accompany aging
and the slowing of dynamics has not yet been seen.

In this paper we study the structure of an aging colloidal glass
by considering how colloids pack together.  Entropy can
be maximized by optimizing packing in a dense suspension.
Consider the intriguing case of the
crystallization of hard spheres.  When spheres arrange into
a crystalline lattice, they lose configurational entropy.
However, they each have more local room to move close to their
lattice site, and thus the vibrational entropy is larger.
This increase in vibrational entropy outweighs the loss of
configurational entropy due to crystallization \cite{hoover68}.  In practice,
this argument holds true for systems with volume fractions
above $\phi_{\rm freeze} = 0.494$, the point at which the
system begins to nucleate crystals; below $\phi_{\rm freeze}$,
the configurational entropy dominates and the system prefers
an amorphous, liquid configuration \cite{pusey86}.

For glasses, we consider a different sort of packing.  The most
efficient way to pack four spheres of diameter $d$ in three
dimensions is to place them at the vertices of a regular
tetrahedron with edge length $d$. In this configuration,
the effective volume fraction for the four spheres reaches a
surprising $0.78$.  In other words, for a given volume fraction
$\phi$, locally four particles can maximize their
entropy by arranging into a regular tetrahedron consistent with
the global volume fraction $\phi$, thus giving them additional
room to move.  However, regular tetrahedra do not tile 3D space
and therefore the most efficient macroscopic packing is that of
a hexagonally packed crystal at $\phi_{\rm hcp}\approx 0.74$.
Thus in a glass there is a {\em frustration} between the
drive to locally pack in tetrahedra to maximize the local
volume available to vibrations, and the inability to tile
3D space with such structures.  This has been suggested as
a possible origin for the glass transition in simple liquids
\cite{nelson84,nelson02,stillinger88,kivelson94}.

We take advantage of the insight afforded
to us by fast laser scanning confocal microscopy
\cite{blaaderen95,kegel00,weeks00,dinsmore01} and study an aging
colloidal glass in terms of tetrahedral packing.  
We focus on geometrical properties of the tetrahedra formed
by the colloids and look for correlations between these {\em
static} quantities and the conspicuous slowing of the {\em
dynamics} as measure by the average tetrahedral mobility. We
find that while the distribution of these static quantities
does not age, they correlate weakly with mobility, suggesting
that the structure facilitates the aging process.

\section{Experimental Methods}

We suspend poly(methyl methacrylate) (PMMA) colloids of
diameter $d=2.36\mu m$ in a mixture of $15\%$ decalin and
$85\%$ cyclohexylbromide by weight. The mixture closely
matches the density and refractive index of the particles,
thus greatly reducing sedimentation and scattering effects. The
size polydispersity of the colloids ($\approx 5\%$) prevents
crystallization.  The particles are sterically stabilized
against van der Waals attractions by a thin layer of
poly-12-hydroxystearic acid \cite{antl}.  We dye the colloids
with rhodamine 6G \cite{dinsmore01}. The particles also carry
a slight charge due to the dye.  In this paper, we measure
all lengths in terms of the diameter $d$ and all times in
terms of $\tau_{\rm diff}$, the time a particle would take to
diffuse its own diameter in the {\em dilute} limit. Given the
solvent viscosity $\eta=2.25$~mPa$\cdot$s at $T=295$~K, this time
is $\frac{d^{2}}{6D}=11.4$~s where $D=\frac{k_{\rm B}T}{3\pi
\eta d}$.

We acquire three dimensional images by fast laser scanning
confocal microscopy at a rate of 1 every 26~s. The observation
volume measures 26$d$ $\times$ 25$d$ $\times$ 4.2$d$. At
these high densities ($\phi>0.58$) the colloids move slowly
and can easily be tracked using established analysis techniques
\cite{weeks00,dinsmore01,crocker96}. The 3D positions of  $\sim$2500
particles are measured virtually instantaneously
with an accuracy of 0.013$d$ in the x-y plane and 0.021$d$
along the optical axis.
We acquire data at least 25$d$ away from the closest wall, to
avoid boundary effects \cite{kose76,gast86}.  We do not observe
any crystals in the bulk even after several weeks.

The phase behavior of this quasi-hard sphere system is
controlled by varying the packing fraction $\phi$. The system
undergoes a glass transition when $\phi>\phi_{g}\approx
0.58$ in agreement with what is seen in hard sphere systems
\cite{pusey86,weeks00}. Here we present data from a sample at
$\phi\approx 0.62$ though we see qualitatively similar results
for all $\phi > \phi_{g}$.

Proper sample initialization is paramount when studying aging
and is ensured here by a vigorous, macroscopic stirring. This
shear melting effectively rejuvenates the glass and yields
reproducible dynamics that depend exclusively on $t_{\rm w}$,
the time elapsed since initialization.  Data acquisition starts
immediately after rejuvenation. Transient macroscopic flows are
observable for the first 25~min$\approx 140 \tau_{\rm diff}$
and we set $t_{\rm w}=0$, or age zero, when they subside. The
results below are insensitive to small variations in this
choice.

\section{Results}

\begin{figure}[t]
\centering
\includegraphics[height=5.5cm]{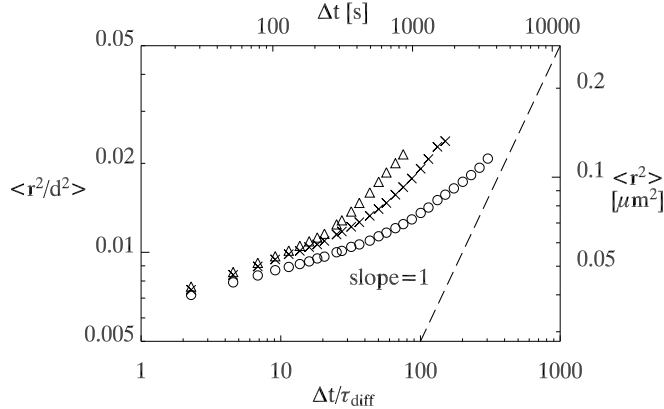}
\caption{Aging mean squared displacement for a
colloidal glass at $\phi\approx0.62$. The three curves
represent three different ages of the sample. $\triangle:
t_{\rm w}=0\tau_{\rm diff}$, $\times:t_{\rm w}=100\tau_{\rm diff}$
and $\bigcirc:t_{\rm w}=300\tau_{\rm diff}$. The dashed line has
a slope of 1 and represents diffusive behavior, not seen in this
glassy sample.}
\label{msd}
\end{figure}

We observe our sample for $\sim 700\tau_{\rm diff}$ without
disturbing it. We  then split the data in three time windows as
follows: $[0-100\tau_{\rm diff}]$, $[100-300\tau_{\rm diff}]$
and $[300-700\tau_{\rm diff}]$. This corresponds to doing
three experiments with samples aged $t_{\rm w}=0$, $100$ and
$300\tau_{\rm diff}$ respectively.  The dynamics slow as the
sample ages, as shown in Fig.~\ref{msd}, where we plot the mean
square displacement for the three data portions averaged over
all particles and over all initial times within a given time
window. At short and medium times ($\frac{\Delta t}{\tau_{\rm
diff}}<10$), particle motions are subdiffusive as indicated
by a slope less than unity on the log-log plot. At longer
times the slope tends to one; the time scale for this upturn
changes dramatically for different values of age $t_{\rm w}$
clearly indicating that the sample is out of equilibrium.
It is this slowing down of dynamics that we wish to analyze
in terms of tetrahedral structure.

% FIGURE ONE -- MSD

\begin{figure}[b]
\centering
\includegraphics[height=5.5cm]{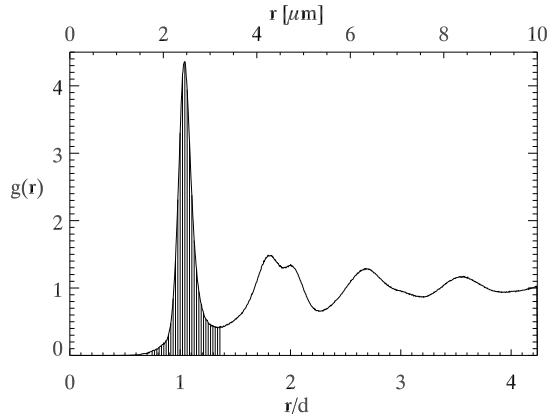}
\caption{Pair correlation function $g(r)$. The shaded
area indicates the range of interparticle distances used
to define nearest neighbors.}
\label{gr}
\end{figure}

We start our structural analysis by calculating the
pair correlation function $g(r)$ and plot the result in
Fig.~\ref{gr}. This function does not vary with age and thus
is calculated by averaging over all times. The first peak
of $g(r)$ is at $r=1.04d$ which deviates somewhat from the
ideal hard-sphere position ($r=d$). This can be explained
by the slight charging mentioned above and perhaps also by
the uncertainty in the value of the particle diameter which
we deem to be at most 2\%. Figure \ref{gr} also shows the
characteristic double second peak found in many glassy systems.

% FIGURE TWO -- gr

In order to study the tetrahedral packing in our sample
we begin by labeling as nearest neighbors every pair of
colloids whose separation is within the first peak of
$g(r)$, namely $0.74d>r>1.38d$, as is shown by the shaded
area in Fig.~\ref{gr}. The lower limit is chosen to eliminate
artificially close pairs which arise from the occasional error
in particle identification, while the upper limit corresponds
to the first minimum of $g(r)$.  Note that a completely coplanar
arrangement of four spheres in a square is excluded as a
tetrahedron, as the diagonal would have length $\sqrt{2}d$ which
is excluded by our upper limit.
The results presented here
are insensitive to small variations in these parameters and
match those obtained using Delaunay triangulation as a nearest
neighbor finding algorithm.

%   The structure of a tetrahedron is uniquely defined by the
%   $(x,y,z)$ coordinates of its vertices, in other words, 12
%   numbers (only 6 of which are independent quantities, given an
%   assumed invariance with translation and rotation).

A tetrahedron is then defined as a quadruplet of mutually
nearest neighbor colloids.  To characterize each tetrahedron,
we calculate several geometrical characteristics.  The first is
``looseness'' $b$, defined as the average of the lengths of the
6 edges, or ``bond lengths'', $b_{\rm i}$.  An ``irregularity''
$\sigma_{b}$ is defined as the standard deviation of the
$b_{\rm i}$.  The looseness and irregularity appear to be
the most important geometric parameters to characterize a
tetrahedron shape, as will be discussed below.
The nondimensional volume $V/d^3$ and nondimensional surface
area $A/d^2$ are also measured.  To quantify an effective
aspect ratio of each tetrahedron, we calculate the height
of the tetrahedron as measured from each of the four faces,
and consider the largest height $H$ and shortest height $h$.
We form aspect ratios from these two heights by dividing by the
areas ($A_{H}$ and $A_{h}$ respectively) of the tetrahedron
face they are perpendicular to.  Conceptually, large values
of $H^2/A_H$ correspond to thin pointy tetrahedra, and small
values of $h^2/A_h$ correspond to flat pancake-like tetrahedra.
We thus term these two quantities ``sharpness'' and ``flatness''
respectively.

In addition to these structural characteristics, we also consider
the dynamics by the
tetrahedral mobility $\mu$, which is calculated by averaging the
distances moved by the four colloids over a time $\Delta
t=50\tau_{\rm diff}$:
\begin{equation}
\mu(t) = {1 \over 4} \sum_{i=1}^4 | \Delta \vec{r_i}(t,\Delta t) | 
\end{equation}
The results that follow do not depend sensitively to the choice
of $\Delta t$.

To assess the value of these structural and dynamical
characteristics, 
we calculate the correlation coefficients
between $\mu$ and the other tetrahedral characteristics
and show them in Table~\ref{cor}.  This is done in the standard
way of defining correlation coefficients,
\begin{equation}
C_{pq} = {1 \over N} \sum_{i=1}^N {(p_i - \bar{p}) \over
\sigma_p} {(q_i - \bar{q}) \over \sigma_q},
\end{equation}
where $p$ and $q$ are any two variables with averages
$\bar{p}$ and $\bar{q}$, and standard deviations $\sigma_p$
and $\sigma_q$.  In our case the sum runs over all tetrahedra
and all times.  In Table~\ref{cor}, a value of one would
signify perfect correlation, a value of -1 would represent
perfect anti-correlation while a value of zero would indicate
completely uncorrelated data.

\begin{table}[tb]
\centering
\begin{tabular}{|l|l|l|l|l|l|l|l|}
	\hline
   & $\mu/d$ & $b/d$ & $\sigma_{b}/d$ & $V/d^{3}$ & $A/d^{2}$ & $\frac{H^{2}}{A_{H}}$ & $\frac{h^{2}}{A_{h}}$  \\
	\hline
 mobility $\mu/d$  & 1 & 0.045 & 0.068 & 0.018 & 0.036 & 0.016 & -0.068 \\
	\hline
 ``looseness'' $b/d$  & - & 1 & 0.37 & 0.82 & 0.98 & -0.30 & -0.50 \\
	\hline
  ``irregularity'' $\sigma_{b}/d$ & - &- & 1 & -0.10 & 0.18 & 0.17 & -0.82 \\
	\hline
 volume $V/d^{3}$ & - & - & -  & 1 & 0.91 & -0.10  & -0.11 \\
	\hline
  surface area $A/d^{2}$ &-  & -  & -  & - & 1 & -0.29 & -0.37 \\
	\hline
  ``sharpness'' $\frac{H^{2}}{A_{H}}$ & - & - & - & -  & - & 1 & -0.090 \\
	\hline
   ``flatness'' $\frac{h^{2}}{A_{h}}$ & - & - & - & -  & - & - & 1  \\
	\hline 
\end{tabular}
\caption{Correlation matrix for some geometrical characteristics
of tetrahedra and mobility. The matrix is symmetric with
respect to the diagonal so the lower half is not repeated.
$b$ is the average length of the tetrahedra edges (``bonds'')
and $\sigma_b$ is the standard deviation of these lengths.
See text for details of the other characteristics.}
\label{cor}
\end{table}

Given that we are trying to understand the slowing of the
dynamics seen in Fig.~\ref{msd}, we focus on the correlation
between mobility $\mu/d$ and the structural characteristics.
While all the coefficients are quite small, those that
relate mobility to looseness $b/d$ and irregularity
$\sigma_b/d$ are relatively big.  Mobility is also noticeably
anticorrelated with the flatness $h^2/A_h$.  Some
insight into these correlations is gained by considering the
correlations between the different structural characteristics,
as shown by the other entries in Table~\ref{cor}.  The
flatness $h^2/A_h$ is strongly anticorrelated with
irregularity $\sigma_b$, and given the more intuitive
nature of $\sigma_b$ and its simpler mathematical definition,
in what follows we focus on $\sigma_b$ rather than $h^2/A_h$.
The volume and area parameters, $V/d^3$ and $A/d^2$, are
strongly correlated with the looseness, which is sensible
given that they all measure the size of a tetrahedron.

We therefore choose to study the looseness and
irregularity as being both relatively well-correlated with
mobility, both easily defined in terms of the six tetrahedron
edge lengths, and weakly correlated with each other (as seen
in Table~\ref{cor}).  The last point suggests that they capture
two distinct properties of tetrahedron structure which are
both important for mobility.

\begin{figure}
%\begin{minipage}[t]{0.5\linewidth} % A minipage that covers half the page
\centering
\includegraphics[height=5cm]{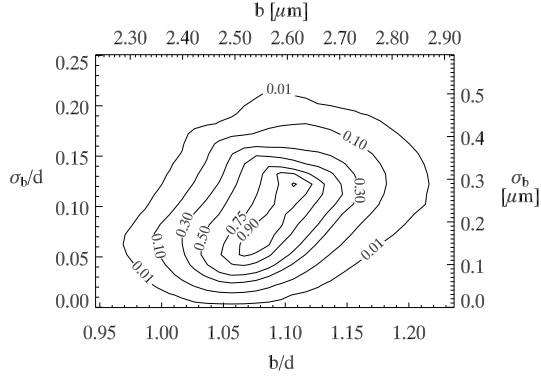}
\caption{Contour plot showing the abundance of tetrahedra with a given looseness and irregularity. The iso-curves are labeled relative to the peak tetrahedral abundance at \{$b/d=1.11; \sigma_{b}/d=0.12$\}.}
\label{count}
%\end{minipage}
%\hspace{0.5cm} % To get a little bit of space between the figures
%\begin{minipage}[t]{0.5\linewidth}
\end{figure}

Figure \ref{count} shows the distribution of tetrahedra in
the $b, \sigma_{b}$-plane. The closed curves represent the
levels of abundance of tetrahedra with a given value of $b$
and $\sigma_{b}$ with respect to the abundance of the most
probable tetrahedron at $b\approx 1.11d$ and $\sigma_b \approx
0.12d$. Somewhat surprisingly, the distribution does not
age \cite{japan05} and we therefore take an average over
all times.  Figure \ref{count} shows a broad variability of
both looseness and irregularity.  However, these curves do
outline a major axis along which many of the tetrahedra lie.
This axis suggests that the looser the tetrahedron the more irregular
it is bound to be.  This is reflected in the correlation
coefficient of $b$ and $\sigma_{b}$ in Table.~\ref{cor},
although its relatively small value (0.37) highlights the breadth of
the overall distribution.

% FIGURE THREE -- count

% FIGURE FOUR -- 2dmob
\begin{figure}
\centering
\includegraphics[height=5cm]{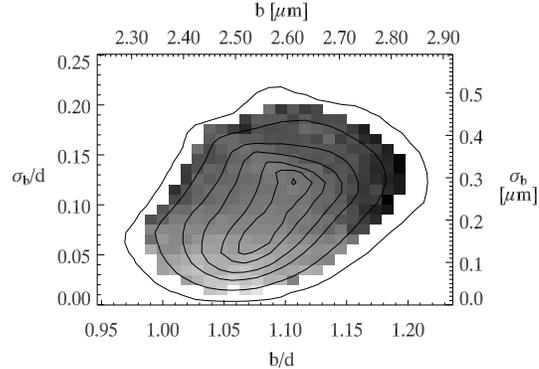}
\caption{Plot of tetrahedral mobility $\langle
\mu(\Delta t=50\tau_{\rm diff})\rangle$, averaged over all ages,
versus looseness $b/d$ and irregularity $\sigma_{b}/d$. The
darker the color the more mobile the tetrahedron. The contour
lines are the same as in Fig.~\ref{count} and represent
the abundance of tetrahedra with a given value of $b/d$
and $\sigma_{b}/d$.}
\label{2dmob}
%\end{minipage}
\end{figure}

We present a qualitative picture of the relationship between
the above two {\em static} geometrical quantities and the
{\em dynamic} quantity of mobility in Fig.~\ref{2dmob} which
shows the average value of mobility as a shade of grey. The
more mobile combinations of $b$ and $\sigma_{b}$ are darker
and the less mobile combinations are lighter. This rendering
gives a clear qualitative view of the correlations between these
quantities. Specifically, we note that mobility increases with
both looseness and irregularity of the tetrahedron.  This makes
intuitive sense:  a larger value of $b$ suggests a smaller local
volume fraction and thus more room to move, and a larger value
of $\sigma_b$ likewise suggests a poorly-packed structure which
may have more room to move.  This also agrees with previous
results seen in supercooled colloidal fluids
\cite{weeks02,harrowell99,conrad05}.
Overplotted on the intensity plot of Fig.~\ref{2dmob} are the
same abundance contours as seen in Fig.~\ref{count}, showing
us that the most probable tetrahedra are a medium shade
of grey - they are neither the fastest or slowest tetrahedra.

% FIGURE FIVE -- mob_vs_b
\begin{figure}[tb]
\centering
\includegraphics[height=5.5cm]{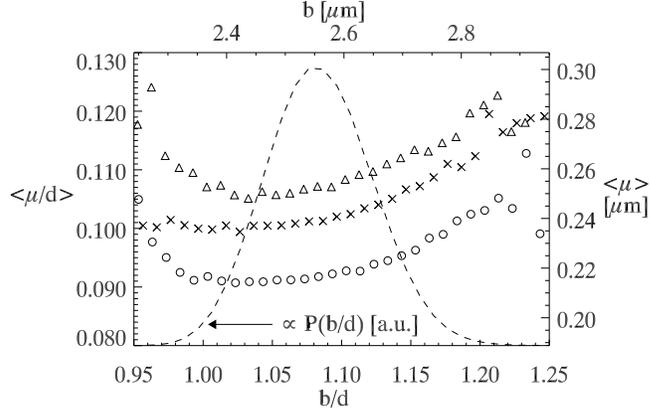}
\caption{Average tetrahedral mobility as a function of looseness $b/d$. The three curves
represent three different ages of the sample. $\triangle:
t_{\rm w}=0\tau_{\rm diff}$, $\times:t_{\rm w}=100\tau_{\rm diff}$
and $\bigcirc:t_{\rm w}=300\tau_{\rm diff}$. The dashed curve represents the distribution $P(b/d)$ and is shown here to highlight the lack of statistics at low ($b/d<0.98$) and high ($b/d>1.2$) values of $b/d$.}
\label{mob_vs_b}
\end{figure}

We thus have two results, an overall slowing of dynamics seen in
Fig.~\ref{msd}, and a relationship between structure and dynamics
seen in Fig.~\ref{2dmob}.  This suggests a possible hypothesis
for aging, that the slowing of the dynamics is an accumulation of
structure corresponding to slower dynamics:  a buildup of
tetrahedra with small values for looseness and irregularity.  As
mentioned earlier, though, the overall distribution of tetrahedra
structural properties (Fig.~\ref{count}) does not depend on the
age of the sample \cite{japan05}.  To reconcile this, we consider in more detail
the connection between structure, dynamics, and the age of the
sample.

We show the influence of looseness on the tetrahedron mobility
in Fig.~\ref{mob_vs_b} where we plot the average mobility of
tetrahedra as a function of looseness.  We do so averaging
over the three sample ages separately. If we consider each
curve separately, we note a reproducible trend:  the least
mobile tetrahedra are those with $b\approx d$ indicating that
those tetrahedra are very tightly packed. At very low values of
$b$ mobility increases somewhat, although as mentioned above,
tetrahedra with $b/d<1$ may be erroneous.   Furthermore, note
that there are extremely few tetrahedra with $b/d<1$ as shown
by the dashed curve representing $P(b/d)$, the probability of
finding a tetrahedron with a given looseness averaged over all
ages. We therefore cannot put too much weight on the values
of $\mu$ for $b<d$. (The same can be said for tetrahedra with
$b>1.17d$.) In the intermediate range there is a clear trend
that looser tetrahedra are more mobile.  Thus the structure
in some way facilitates the aging, in that looser regions
are more free to rearrange.  However, it is also important to
note that each symbol is an average over a broad distribution
of mobilities associated with the given value of $b/d$. In
particular the standard deviation of the distribution is almost
comparable with the average value. This simply means that the
correlation between $b$ and $\mu$ is a weak, average effect and
not, for example, a usefully predictive relationship
\cite{conrad05}.

As expected with any plot involving the mobility of this system,
aging is clearly visible in Fig.~\ref{mob_vs_b} as the three
curves are shifted down to lower mobilities as $t_{\rm w}$
increases. The overall shape of the curves, however, does
not depend on the age of the system.  In other words, we are
not witnessing a relative shift in the mobility of tetrahedra
with varying looseness but merely an overall slowing down of
all tetrahedra.

% FIGURE SIX -- mob_vs_sigma
\begin{figure}[tbh]
\centering
\includegraphics[height=5.5cm]{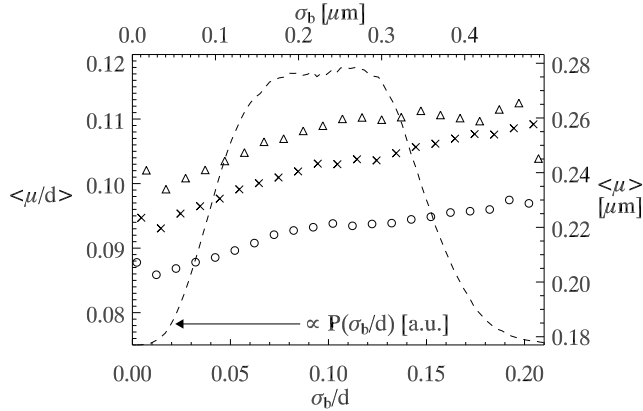}
\caption{Average tetrahedral mobility as a function of
irregularity $\sigma_b/d$. The three curves represent
three different ages of the sample. $\triangle: t_{\rm
w}=0\tau_{\rm diff}$, $\times:t_{\rm w}=100\tau_{\rm diff}$
and $\bigcirc:t_{\rm w}=300\tau_{\rm diff}$. The dashed curve
represents the distribution $P(\sigma_{b}/d)$ and is shown here
to highlight the lack of statistics at low ($\sigma_{b}/d<0.01$)
and high ($\sigma_{b}/d>0.2$) values of $\sigma_{b}/d$.}
\label{mob_vs_sigma}
\end{figure}

A similar analysis on the relation between tetrahedral mobility
and irregularity is shown on Fig.~\ref{mob_vs_sigma}. As a
reference we plot $P(\sigma_{b}/d)$, the probability
of finding a tetrehedron with a given irregularity. This
distribution does not age and is therefore averaged over all
times. Again, we look at the dependence of $\mu$ on $\sigma_b$ for the
three ages in our experiments and find that there is a positive
correlation, as previously indicated in Table~\ref{cor}. Just
as in the case of $b$, this positive correlation is an
average effect and again, the distribution that leads to each
of the symbols on the figure is quite broad. Nevertheless,
a reproducible difference of $\sim 10\%$ in the mobility
differentiates very regular tetrahedra from very irregular
ones. Again, just as above, aging is evident in the data and
again it has no strong effect on the shape of the curves but rather
uniformly slows down tetrahedra with all values of irregularity.

\section{Conclusion}

We observe colloidal glasses and find clear signs of aging in
the mean squared displacement of the particles (Fig.~\ref{msd}).
We analyze the static structure of the aging sample in terms
of tetrahedral packing of colloidal particles.  We find a broad
distribution of tetrahedra as measured by the distributions of
tetrahedral ``looseness'' and ``irregularity'', corresponding
to the tetrahedron's mean edge length and the standard
deviation of edge lengths, respectively (Fig.~\ref{count}).
These two quantities are slightly correlated; on average,
the looser a tetrahedron is the more irregular it will be.
More importantly, we find that tetrahedral shape and mobility
are somewhat correlated: the looser and the more irregular a
tetrahedron is the higher its mobility (Fig.~\ref{2dmob}).
This suggests that aging might be due to an increase in
tight, regular tetrahedral structure, but surprisingly the
distribution of geometrical quantities is age-independent.
Instead we find that aging indiscriminately affects tetrahedra
with all values of looseness and irregularity by uniformly
decreasing their mobility.

In conclusion, we find that static structure as measured by
tetrahedral quantities does not indicate the age of a glass.
None of the distributions of the geometrical quantities
considered in Table~\ref{cor} show any aging. However, at any
instant in time the age of our sample must somehow be encoded
in the positions of the colloids and, for example, analyzing the
spatial correlations between tetrahedra, while beyond the scope
of this paper, might provide more insight into this matter.
Finally, since aging ought to result in subtle configuration
changes, and since the looser and more irregular tetrahedra
allow for the most motion to happen, we can infer that the
local structure does indeed facilitate the aging process.
While it is worth noting that it is not established that the
most mobile particles are the most important ones for aging,
the connection between structure and mobility holds true
for less mobile particles as well.  Table~\ref{cor} suggests
that in this respect, tetrahedral irregularity is the most
significant quantity whose positive correlation with mobility
lends support to our original motivating idea that perfect
tetrahedra are an important structural element in a glass.

% The Appendices part is started with the command \appendix;
% appendix sections are then done as normal sections
% \appendix

% \section{}
% \label{}

\begin{ack}
We thank T.~Brzinski, P.~Harrowell, and C.~Nugent for useful discussions.
This work was supported by NASA microgravity fluid physics grant NAG3-2728.
\end{ack}

\end{document}